**Some properties of the simple nonlinear recursion $y(\ell+1) = [1 - y(\ell)]^p$ with $p$ an *arbitrary positive* integer**

Francesco Calogero

Physics Department, University of Rome "La Sapienza", Rome, Italy

Istituto Nazionale di Fisica Nucleare, Sezione di Roma 1, Rome, Italy

Istituto Nazionale di Fisica Matematica, Gruppo Nazionale di Fisica Matematica, Italy

francesco.calogero@uniroma1.it, francesco.calogero@roma1.infn.it,

**Abstract**

It is shown that the behavior of the solutions of the *nonlinear* recursion $y(\ell+1) = [1 - y(\ell)]^p$— where the dependent variable $y(\ell)$ is a *real* number, $\ell = 0, 1, 2...$ is the *independent* variable, and $p$ is an *arbitrary positive integer*— is *easily ascertainable*.

**1. Introduction**

In this paper we investigate the simple *recursion*

$$y(\ell+1) = [1 - y(\ell)]^p \ , \tag{1}$$

where $y(\ell)$ is the *dependent* variable, $\ell = 0, 1, 2...$ is the *independent* variable (think of it as, say, a *ticking time*) and $p$ is an *arbitrary positive integer*.

For $p = 1$ the solution of the *initial-value* problem of this recursion

$$y(\ell+1) = 1 - y(\ell) \tag{2a}$$

is *trivially ascertainable*: it is

$$y(\ell) = 1 - y(0) \ \text{ for } \ell = 1, 3, 5, ... \ , \tag{2b}$$
$$y(\ell) = y(0) \ \text{ for } \ell = 2, 4, 6, ... \ ; \tag{2c}$$

so the solution is always *periodic* with period 2. And of course it also features the trivial *equilibrium* solution

$$y(\ell) = 1/2 \ . \tag{2d}$$

For $p = 2$ the recursion (1) was recently investigated in Ref. [1], where it is shown that the behavior of its solutions may be *rather explicitly* described, although— to the best of my knowledge—it cannot be expressed in terms of known functions (maybe these solutions will eventually be given a name, if this recursion turns out to become *very* useful in *applicative* contexts).

We restrict hereafter attention to the case in which the solution $y(\ell)$ is a *real* number (this being of course the main case of possible *applicative* relevance); and we show that this kind of full understanding is also possible for the more general recursion (1) with $p$ an *arbitrary positive integer*. And let us note to begin with that, for *any positive integer* $p$ and *any real initial datum* $y(0)$, clearly the solution $y(\ell)$ of the recursion (1) is—for *all* values of the *independent* variable $\ell$—a (finite) *real number*.

**2. Special solutions**

The simplest solutions of the recursion (1) are those which are in fact $\ell$-*independent*,
$$y(\ell) = \overline{y} \,, \qquad (3a)$$
where $\overline{y}$ must of course then be a (*real!*) solution of the *algebraic* equation
$$\overline{y} = (1-\overline{y})^p \,. \qquad (3b)$$
Note that at least 1 *real* solution of this equation *always* exists *iff* $p$ is an *odd positive integer*.

There are moreover always 2 *periodic* solutions—with period 2—namely those starting from the *initial values* $y(0) = 0$ and $y(0) = 1$:

$$y^{(e)}(\ell) = 1 \text{ if } \ell = 0, 2, 4, \ldots,$$
$$y^{(e)}(\ell) = 0 \text{ if } \ell = 1, 3, 5, \ldots ;$$
$$y^{(o)}(\ell) = 0 \text{ if } \ell = 0, 2, 4, \ldots,$$
$$y^{(o)}(\ell) = 1 \text{ if } \ell = 1, 3, 5, \ldots ;$$
$$y^{(e,o)}(\ell + 2) = y^{(e,o)}(\ell) \,. \qquad (3c)$$

**3. The p = 3 case**

The $p = 2$ case was treated in Ref. [1], so we begin by focusing on the next case, $p = 3$, hence on the recursion
$$y(\ell+1) = [1 - y(\ell)]^3 \,. \qquad (4)$$

Its *equilibria* are of course the roots of the *cubic* equation
$$\overline{y}^3 - 3\overline{y}^2 + 4\overline{y} - 1 = 0 \,; \qquad (5a)$$
this *cubic* equation has only 1 *real* root, namely the *irrational* number
$$\begin{aligned}\overline{y} &= 1 + \left[(\sqrt{93}-9)/2\right]^{1/3} \cdot 3^{-2/3} - \left[(\sqrt{93}-9)/2\right]^{-1/3} \\ &= 0.3176721961719808 \end{aligned} \qquad (5b)$$
(the other 2 roots being 2 *complex* numbers of each other, not reported here).

Note that above and hereafter numbers written in decimal form are only represented approximately.

It is of course plain that the recursion (4) features 2 *periodic* solutions, of period 2, identified above, see (3c).

It is moreover easy to prove that, for *every initial* datum $y(0)$ in the interval $0 < y(0) < 1$—with the *sole* exception of the *equilibrium* value $y(0) = \bar{y}$ (see eq. (5b))—the behavior of *all* these solutions is to *jump at every step*—as $\ell$ increases *step by step* from its *initial* value $\ell = 0$ towards $\ell = \infty$—from one side to the other of the *equilibrium* value $\bar{y}$, getting progressively more and more away from that *equilibrium* value and instead approaching *steadily* (or rather *jumpily*) the 2 boundaries of the interval of $y$-values from 0 to 1; hence all these solutions are *asymptotically periodic* with period 2 as $\ell \to \infty$. For instance if $y(0) = 1/2$, the recursion (4) yields

$$\begin{aligned}
y(1) &= 1/8 = 0.125\ , \\
y(2) &= (7/8)^3 = 343/512 = 0.66992186\ , \\
y(3) &= \left[1 - (7/8)^3\right]^3 = (169/512)^3 = 0.03596252948\ , \\
y(4) &= \left[1 - (169/512)^3\right]^3 = 0.89594581\ .
\end{aligned} \qquad (5c)$$

Note that this description provides a quite neat—if not numerically precise—yet completely cogent description of *all* the solutions of the recursion (4), for any *initial* datum $y(0)$ in the interval $0 < y(0) < 1$.

And it is likewise easy to show that *all* solutions starting instead *outside* of that interval $0 < y(0) < 1$ shall display a, to some extent, somewhat analogous yet quite different behavior, by *jumping* at every step from one side of the *interval* $0 < y < 1$ to the *other* side and thereby *increasing* steadily (at every *second* step) in *modulus*, and eventually *diverging asymptotically* (towards $\pm\infty$) as $\ell \to \infty$. For instance if $y(0) = 2$, then $y(1) = -1$, $y(2) = (2)^3 = 8$, $y(3) = (-7)^3 = -343$, ... .

It is thus seen that a rather detailed description of the behavior of *all* the solutions of the recursion (4) is available.

## 4. The p = 4 case

In this Section we discuss the solution of the recursion (1) in the $p = 4$ case,

$$y(\ell + 1) = [1 - y(\ell)]^4\ . \qquad (6)$$

Note that in this case, whatever the value is of the *initial datum* $y(0)$, for *all* subsequent values of the *ticking time* $\ell = 1, 2, 3...$, the dependent variable $y(\ell)$ shall be *positive*, $y(\ell) > 0$ (except for the special case with $y(0) = 1$, see (3c)). In this respect the situation is quite *analogous* to the case discussed in Ref. [1], as well of course as in *all* cases when the exponent $p$ in the recursion (1) is an *even* positive *integer* (see below). So in this **Section 3** we restrict our consideration only to *positive* values of the dependent variable $y(\ell)$.

Obviously this recursion (6) also features the 2 *periodic trivial* solutions (3c).

It is also *obvious* that, if the solution $y(\ell)$ starts inside the interval $0 < y(0) < 1$, it shall remain *inside* that interval thereafter, hence $0 < y(\ell) < 1$. But how shall it evolve within that interval?

Then the first question to address is the possible presence of *equilibria inside* that interval, namely of *real* solutions of the *algebraic fourth-order* equation

$$\bar{y} = (1 - \bar{y})^4 \ , \tag{7a}$$

clearly identifying the possible *equilibria* of the recursion (1) with $p = 4$. Of course this algebraic equation has only 4 roots. With some help from **Mathematica** we display these 4 roots, only 2 of which are *real* numbers, the other 2 being instead 2 *complex-conjugate* numbers of each other (they are listed below in their *approximate decimal* versions):

$$\begin{align} \bar{y}^{(1)} &= 0.275508 \ , \\ \bar{y}^{(2)} &= 2.22074 \ , \\ \bar{y}^{(3,4)} &= 0.751874 \pm 1.03398 \cdot \mathbf{i} \ , \tag{7b} \end{align}$$

where clearly $\mathbf{i}$ is the *imaginary unit* ($\mathbf{i}=\sqrt{-1}$). Note that *only one* of the 2 *real* roots falls *inside* the interval $0 < y < 1$.

Again, it is then easily seen that *any* solution which starts *inside* the interval from 0 to 1 (so that $0 < y(0) < 1$)—excluding *only* the *equilibrium* one sitting *forever* at $y(\ell) = \bar{y}^{(1)}$—shall then evolve from its *initial* value $y(0)$ by jumping at every subsequent step *over* the *equilibrium* value $\bar{y}^{(1)}$ (see the eqs. (7b)); performing, as it were, a *back-and-forward* dance—but always remaining *inside* the interval $0 < y(\ell) < 1$ and indeed at every step getting *closer* to its 2 border-values 0 and 1. So *all* these solutions $y(\ell)$ shall be *asymptotically isochonous* with period 2, getting alternatively closer and closer to one and then to the other of the 2 border-values, 0 and 1, at every subsequent step of the ticking-time independent variable $\ell$, as it increases without limit, $\ell = 1, 2, , 3, ..., \infty$.

For instance, if $y(0) = 1/2$, then clearly $y(1) = 1/16$, $y(2) = (15/16)^4$, $y(3) = \left[1 - (15/16)^4\right]^4$, and so on.

There remains to investigate what the behavior shall be if the *initial* datum $y(0)$ is *outside* the interval $0 < y(\ell) < 1$.

The first—obvious and relevant—remark is that, if $y(0)$ is in the interval $1 < y(0) < 2$, clearly $y(1)$ shall be in the interval $0 < y(1) < 1$, hence its further evolutions shall be as described a few lines above. And clearly the same shall be true (at least) for all values of $y(0)$ up to $\bar{y}^{(2)}$ (see the eqs. (7b)).

What remains to be ascertained is what happens if $y(0) > \bar{y}^{(2)}$. So let us investigate what happens if $y(0) = \bar{y}^{(2)} + \varepsilon$ with $\varepsilon$ *positive* and *infinitesimal*. Then

$$\begin{align} y(1) &= \left(\bar{y}^{(2)} - 1 + \varepsilon\right)^4 = \left(\bar{y}^{(2)} - 1\right)^4 + 4 \cdot \left(\bar{y}^{(2)} - 1\right)^3 \varepsilon \\ &= \bar{y}^{(2)} + 4 \cdot (1.22074)^3 \varepsilon > y(1) \tag{7c} \end{align}$$

and this clearly also entails that thereafter $y(\ell)$ shall steadily increase, *diverging* to $+\infty$ as $\ell \to \infty$. And clearly the same outcome shall obtain from *all initial* values $y(0) > \bar{y}^{(2)}$. For instance if $y(0) = 3$, then $y(1) = 8$, $y(2) = 7^4 = 2401$, $y(3) = (2400)^4$, and so on.

It is thus seen that a rather detailed description of the behavior of *all* the solutions of the recursion (6) is available.

### 5. The cases with p odd and larger than 4

It is easy to convince oneself that the phenomenology in *all* these cases is *quite analogous* to that described in **Section 3**, with the *single real root* of the relevant *algebraic* equations of *odd* degree $p > 4$

$$\overline{y} = (1 - \overline{y})^p \tag{8}$$

getting progressively *smaller* as $p$ increases: for instance for $p = 5, 7, 9, 11$ the (approximate) values of those *single* real roots $\overline{y}$ are respectively 0.243, 0.203, 0.176, 0.156.

### 6. The cases with p even and larger than 5

Again, the phenomenology in *all* these cases is *quite analogous* to that described in **Section 4**, with the *only* 2 *real roots* of the relevant *algebraic* equations of *even* degree $p > 5$

$$\overline{y} = (1 - \overline{y})^p \tag{9}$$

being given—for the following *even* values of $p = 6, 8, 10, 12$—by the following *couples* $\overline{y}^{(1)}$, $\overline{y}^{(2)}$ of (approximate) values: 0.222, 2.13; 0.188, 2.10; 0.165, 2.08; 0.147, 2.06.

### 7. Outlook

The results reported in this paper are rather *elementary* mathematical findings; their potential usefulness is presumably mainly in *applicative* contexts.

Obviously an easy extension of the findings described above obtains via the simple change of dependent variable

$$x(\ell) = a \cdot y(\ell) + b \tag{10}$$

where $a$ and $b$ are 2 *a priori arbitrary real* numbers and $x(\ell)$ is the *new dependent variable*.

More generally, the findings reported in this paper might be utilized to investigate *systems of recursions*, in analogy to what was done in the 3 papers [2-4] by utilizing the findings reported in [1]. And analogous possible generalizations involve the replacement of the *dependent* variable with more general mathematical entities, like *vectors*, *matrices*, *operators*,...; as well as the extension to broader categories of *independent* variables than just *positive integers*, including moreover the possibility that there be more than just 1 *independent* variable. Of course not all these developments may be relevant in *applicative* contexts; but some shall.

I might make myself some progress in these directions in the future, but due to my age any such development shall be *quite slow*. Other scientists are of course most welcome to work in such directions; I will be most grateful to those who might do so, especially if they will promptly update me by *e-mail* about their findings.